\documentclass[english,twocolumn]{revtex4}
\usepackage[T1]{fontenc}
\usepackage[latin9]{inputenc}
\usepackage{bm}
\usepackage{graphicx}
\usepackage{amssymb}

\usepackage{babel}

\begin{document}

\title{Decoherence in Weakly Coupled Excitonic Complexes}

\author{Tomá\v{s} Man\v{c}al$^{1}$, Vytautas Balevi\v{c}ius Jr.$^{1,2}$,
and Leonas Valkunas$^{2,3}$ }

\affiliation{$^{1}$Charles University in Prague, Faculty of Mathematics and Physics,
Ke Karlovu 5, CZ-121 16 Prague 2, Czech Republic, $^{2}$Department
of Theoretical Physics, Faculty of Physics of Vilnius University,
Sauletekio Avenue 9, build. 3, LV-10222 Vilnius and $^{3}$Institute
of Physics, Savanoriu Avenue 231, LV-02300 Vilnus, Lithuania }
\begin{abstract}
Equations of motion for weakly coupled excitonic complexes are derived.
The description allows to treat the system in the basis of electronic
states localized on individual chromophores, while at the same time
accounting for experimentally observable delocalization effects in
optical spectra. The equations are show to be related to the well-known
Förster type energy transfer rate equations, but unlike Förster equations,
they provide a description of the decoherence processes leading to
suppression of the resonance coupling by bath fluctuations. Linear
absorption and two-dimensional photon echo correlation spectra are
calculated for simple model systems in homogeneous limit demonstrating
distinct delocalization effect and reduction of the resonance coupling
due to interaction with the bath.
\end{abstract}
\maketitle

\section{Introduction}

Excitonic interaction determines spectroscopic and functional properties
of many naturally occurring, as well as artificially synthesized macromolecular
systems \cite{Agranovich82,Pope99,PhExcitons,PolymerExcitons,Silinsh}.
The great variability of photosynthetic antennae of plants and bacteria
that involve only a limited number of different types of small molecules
as building blocks, is to a high degree enabled by the large influence
of inter-chromophore interactions on spectral and energy transfer
properties of closely packed protein-chromophore complexes \cite{BlankenshipBook,PhExcitons}.
In recent years, some of the properties and function of photosynthetic
antennae were linked both theoretically \cite{Cheng2006,Mancal06a}
and experimentally \cite{Engel07a,LeeScience,ColliniScience} to various
types of coherence effects. In some photosynthetic complexes, dynamics
related to long-lasting electronic coherences was observed during
relaxation of excitation energy at $77$ K and even at room temperature
\cite{Engel07a,Panitchayangkoon10}, and it was speculated that the
coherent mode of energy transfer can improve the robustness of the
energy transfer process \cite{Engel07a,Cheng-Review}. While quantitative
aspect of this improvement is a matter of ongoing research, another
closely related type of coherence, the one accompanying delocalization
of excited states in weakly coupled chromophore complexes, has been
found to play a significant role in spectra and energy relaxation
rates of some bacterial antenae. For the peripheral bacterial light
harvesting complexes LH2 and LH3, it has been shown that event at
very weak resonance coupling the absorption spectra, relaxation rates
and coherent two-dimensional electronic spectra show characteristic
features of delocalization \cite{Cheng2006,Zigmantas06a}. 

Proteins fix the positions of chromophores in a protein-chromophore
complex, and thus determine their mutual interactions. However, the
protein surrounding also tune local excitation energies of the chromophores
to achieve further flexibility of the antennae \cite{RengerFMO}.
The protein environment influences the electronic degrees of freedom
(DOF) also on the ultrafast time scale. The shape of the absorption
line of a separate molecule is defined by the interaction of electronic
transition with the intramolecular vibrations and vibrations (phonons)
of a molecular surrounding. The concept of line-shape function originates
from pioneering theoretical work by Lax on absorption spectrum of
a two-level system coupled to harmonic oscillators \cite{Lax52}.
Further development of the theory has been based on the stochastic
approach \cite{Anderson53,Kubo62}, description of anharmonicity \cite{Georgievskii99}
and the Brownian oscillator model for the phonon system \cite{MukamelBook}.
In molecular aggregates the intermolecular interactions compete with
the electronic transition coupling to the intramolecular vibrations
and phonons, which both are usually considered as thermal bath fluctuations. 

\begin{figure}
\includegraphics[clip,width=1\columnwidth]{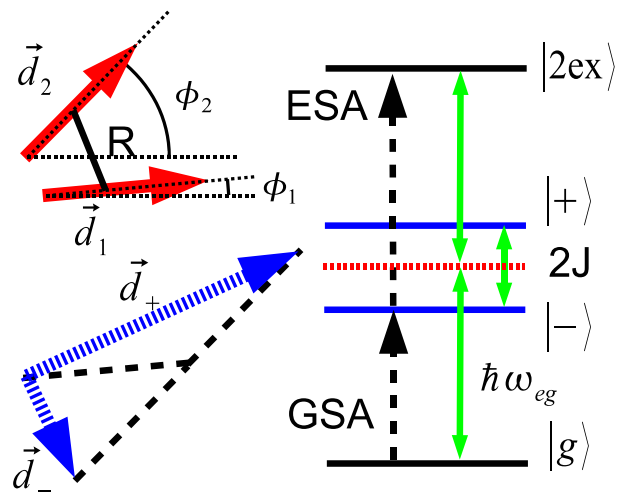}

\caption{\label{fig:illustration}Illustration of the excitonic effect in a
homodimer. Electronic excited states $|1\rangle$ and $|2\rangle$
with transition energy $\hbar\omega_{eg}$ localized on the individual
molecules (depicted by their transition dipole vectors $\bm{d}_{1}$
and $\bm{d}_{2}$) form delocalized eigenstates $|+\rangle$ and $|-\rangle$
of the total electronic Hamiltonian. The result of excitonic interaction
in a homodimer is splitting of transition energies by twice the resonance
interaction $J$, and consequently an offset of ESA with respect to
ground state absorption. The delocalized states have new transition
dipole moments $\bm{d}_{+}$ and $\bm{d}_{-}$ corresponding to a
sum and a difference of $\bm{d}_{1}$ and $\bm{d}_{2}$, respectively. }

\end{figure}

Two-limiting cases of relative strength of the resonance interaction
and the electron-phonon coupling are widely used in various theoretical
methods. In the strong exciton coupling regime (with respect to electron
phonon interaction) the homogeneous bandwidths of the absorption/emission
spectra and spectral dynamics are related to the exciton relaxation
(dephasing) caused by the thermal bath fluctuations. Excitonic splitting
of the transition energies due to the resonance interaction schematically
illustrated in Fig. \ref{fig:illustration}, dominates absorption
spectrum. In the opposite regime, the intermolecular resonance interaction
might be considered perturbatively. The excitation dynamics is then
characterized by hopping between the chromophore molecules, and can
be well described by the Förster theory. Förster resonance energy
transfer (FRET) is nowadays widely accepted as a molecular ruler in
various biological systems \cite{Ha03,Lipman03}. 

The above mentioned photosynthetic complex LH2 is a good example of
a system where both these limiting cases persist. This complex is
arranged as a highly symmetric ring of 9 (or 8 depending on the species
of bacteria) protein-chromophore subunits, each containing two helical
trans-membrane polypeptides, the $\alpha$-polypeptide on the inner
side and the $\beta$-polypeptide on the outer side of the ring \cite{PhExcitons}.
The carboxy-terminal domain of this protein binds, in the hydrophobic
membrane phase, a ring of 18 (or 16 depending on the species of bacteria)
tightly coupled bacteriochlorophyll (Bchl) molecules with a center-to-center
distance of less than 1 nm between neighboring chromophores. This
ring is responsible for the intense absorption of LH2 at 850 nm (the
so called B850 ring). Due to the relatively small distances between
the chromophores in the B850 ring the interaction between Bchl molecules
plays an important role in determining their spectroscopic and functional
properties. A second ring of 9 (or 8) weakly interacting Bchls is
bound by the amino-terminal domain of LH2 (chromophore-chromophore
distance of about 2.1 nm) and is largely responsible for the absorption
at 800 nm (the B800 ring). Most of the spectroscopic results for B850
band can be well explained in terms of the exciton model invoking
available structural data \cite{MecDermott95,Koepke95}. However,
the single molecular fluorescence data and studies of the temperature
dependence of the absorption spectra reveal that the conventional
exciton model has to be modified due to the exciton interaction with
the protein surrounding. The modified (dichotomous) exciton model
was proposed to explain these discrepancies \cite{Valkunas07,Janusonis08,Zarlauskiene08}.
In contrast, the absorption and excitation dynamics in the B800 ring
are usually considered in terms of localized excitations of individual
chromophores, due to the weakness of the intermolecular coupling.
Despite of that, signatures of weak excitonic coupling have also been
identified in the optical spectra of photosynthetic antennae of LH2
\cite{Cheng2006}, single molecular excitation spectra at low temperatures
\cite{vanOijen99}, and 2D spectra of its analogous LH3 complex \cite{Zigmantas06a}.
Similar effects can be expected in other molecular aggregates with
weakly coupled chromophores, DNA stacks, polymers, etc.

In this paper we formulate and develop a theoretical model, which
allows us to describe exciton dynamics in the system of weakly coupled
molecular aggregates in terms of the reduced density matrix (RDM)
in the basis of the site representation of the chromophores. The exciton
coupling is treated as a perturbation and equations of motion (EM)
for the RDM are derived in the second order. This enables us to retain
main excitonic effects such as transition dipole moment redistribution,
transition energy shift and a characteristic excited state absorption
(ESA) shift, while simultaneously working with localized states. 

The paper is organized as follows. In the next section, we introduce
the model system and its Hamiltonian. In Section \ref{sec:Equations-of-Motion}
we derive EM for a general weakly coupled excitonic complex, and in
Section \ref{sub:Long-time-limit} we discuss long time limit of these
equations, leading to Förster type of relaxation. We discuss calculations
of linear absorption and two-dimensional photo-echo spectra in Section
\ref{sec:Two-dimensional-Correlation-Photon}. Model calculations
are presented in Section \ref{sec:Results-and-Discussion}.

\section{Model Hamiltonian\label{sec:Model-Hamiltonian}}

Let us consider an aggregate of $K$ two-level chromophores. The electronic
ground state of such an aggregate can be described by a product state

\begin{equation}
|g\rangle\equiv|g_{1}\dots g_{K}\rangle\equiv|g_{1}\rangle|g_{2}\rangle\dots|g_{K}\rangle,\label{eq:ground}\end{equation}
where the state $|g_{i}\rangle$ is the ground state of the $i-$th
monomeric chromophore. Excited states of the aggregate can be constructed
using monomeric excited states $|e_{i}\rangle$. Thus, we will represent
single- and double-excitation states as

\begin{equation}
|n\rangle\equiv|g_{1}g_{2}\dots e_{n}\dots g_{K}\rangle,\label{eq:1ex}\end{equation}
 \begin{equation}
|N\rangle\equiv|(n,m)\rangle\equiv|g_{1}\dots e_{n}\dots e_{m}\dots g_{K}\rangle,\quad n<m.\label{eq:2ex}\end{equation}
Excited states with more than two excitation will not be considered
as they can often be neglected in the first (linear) and third order
spectroscopic experiments that we have in mind here. Unless stated
otherwise, we will use capital letters to denote a double-excitation
index, e.g. $A=(a,b)$, and lower case letters to denote one-excitation
states. The diad $(a,b)$ will be used to denote two-excitation states
when the knowledge of the underlying one-excitation states is required.
Also sums over double-excitation states will be used $\sum_{A}$ or
$\sum_{(a,b)}$ with the meaning of $\sum_{a=1}^{K}\sum_{b=a+1}^{K}$.

Individual chromophores are described by their ground- and excited
state electronic energies $\epsilon_{i}^{g}$ and $\epsilon_{i}^{e}$,
nuclear potential energy surfaces of the ground- and excited states
$V_{i}^{g}(Q)$ and $V_{i}^{e}(Q)$ and the nuclear kinetic energy
$T_{i}(P)$. If we assume for a while that individual chromophores
in the aggregate do not interact we arrive at the following Hamiltonian

\begin{equation}
H_{0}=H_{B}+H_{el},\label{eq:H0}\end{equation}
 where\[
H_{el}=\epsilon_{g}|g\rangle\langle g|+\sum_{n=1}^{K}\left(\epsilon_{n}+\langle V_{n}(Q)-V_{g}(Q)\rangle\right)|n\rangle\langle n|\]

\begin{equation}
+\sum_{N=1}^{K(K-1)/2}\left(\epsilon_{N}+\langle V_{N}(Q)-V_{g}(Q)\rangle\right)|N\rangle\langle N|,\label{eq:Hel}\end{equation}
 and\[
H_{B}=\left(T(P)+V_{g}(Q)\right)|g\rangle\langle g|\]
\[
+\sum_{n=1}^{K}\left(V_{n}(Q)-\langle V_{n}(Q)-V_{g}(Q)\rangle\right)|n\rangle\langle n|\]
 \begin{equation}
+\sum_{N=1}^{K(K-1)/2}\left(V_{N}(Q)-\langle V_{N}(Q)-V_{g}(Q)\rangle\right)|N\rangle\langle N|.\label{eq:HB}\end{equation}
Here, we introduced \begin{equation}
T(P)=\sum_{i=1}^{K}T_{i}(P),\; V_{g}(Q)=\sum_{i=1}^{K}V_{i}^{g}(Q),\label{eq:Tcelk}\end{equation}
\[
V_{n}(Q)=\sum_{i\neq n}^{K}V_{i}^{g}(Q)+V_{n}^{e}(Q),\; V_{N=(k,l)}(Q)\]
 \begin{equation}
=\sum_{i\neq k,l}^{K}V_{i}^{g}(Q)+V_{k}^{e}(Q)+V_{l}^{e}(Q).\label{eq:Vcelk}\end{equation}
 The mean value of the difference between the potential energy in
the state denoted by index $\alpha$ and potential energy in the electronic
ground state was denoted as\begin{equation}
\langle V_{\alpha}(Q)-V_{g}(Q)\rangle=Tr_{Q}\{(V_{\alpha}(Q)-V_{g}(Q))W_{eq}\},\label{eq:average}\end{equation}
where $\alpha=n,N$. The density operator $W_{eq}$ represents the
equilibrium state of the nuclear DOF in the electronic ground state.

\begin{figure}
\includegraphics{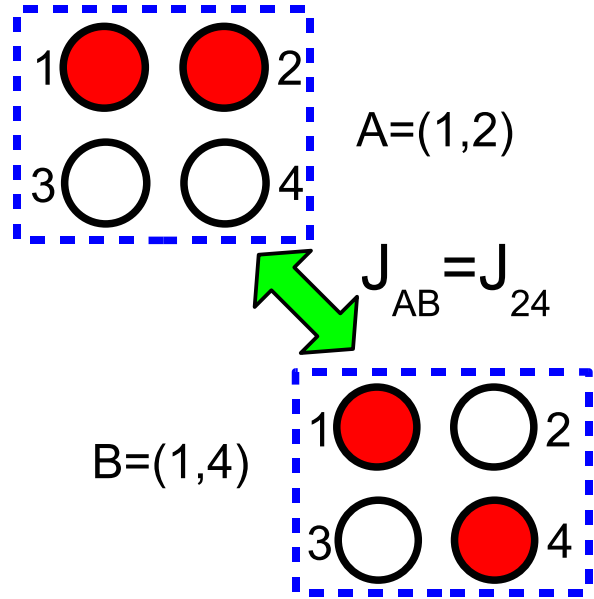}

\caption{\label{fig:two-ex-states}Coupling between two two-excitation states
$A=(1,2)$ and $B=(1,4)$. The coupling $J_{24}$ enables transition
between the two states, while the one-excitation state $1$ is shared.}

\end{figure}

Introducing excitonic interaction, the total Hamiltonian reads

\begin{equation}
H=H_{0}+H_{J},\label{eq:Htot}\end{equation}
 \begin{equation}
H_{J}=\sum_{nm}J_{nm}|n\rangle\langle m|+\sum_{NM}J_{NM}|N\rangle\langle M|.\label{eq:Hj}\end{equation}
 For two distinct double-excitation states $M=(m,n)$ and $N=(k,l)$
we assume the following \emph{ansatz} for the resonance coupling\[
J_{MN}=J_{(m,n)(k,l)}=\delta_{mk}J_{nl}+\delta_{nl}J_{mk}\]
 \begin{equation}
+\delta_{ml}J_{nk}+\delta_{nk}J_{ml}.\label{eq:2ex_coupl}\end{equation}
 The \emph{ansatz} is schematically described in Fig. \ref{fig:two-ex-states}
which demonstrates how two two-exciton states transfer from one to
another by a transition of one excitation, while the other is shared
by both double-excitation states.

It is important to note that due to the resonance coupling Hamiltonian,
$H_{J}$, neither the single- nor double-excitation states are eigenstates
of the total electronic Hamiltonian. The actual eigenstates of the
electronic Hamiltonian (obtained by its diagonalization) can be expressed
as linear combinations of the single- and double-excitation states,
and they are usually termed one- and two-exciton states, respectively.

\section{Equations of Motion\label{sec:Equations-of-Motion}}

\subsection{Projection operator method}

We start investigation of the dynamics of the molecular aggregates
with the Liouville--von Neuman equation for the total density matrix
$W(t)$, \begin{equation}
\frac{\partial}{\partial t}W(t)=-i{\cal L}W(t).\label{eq:Liouville-v-N}\end{equation}
 Here, we defined the Liouville superoperator (or Liouvillian) ${\cal L}A=\frac{1}{\hbar}[H,A]$.
The total Liouvillian can be written in terms of the system and the
resonance interaction Liouvillians \begin{equation}
{\cal L}={\cal L}_{0}+{\cal L}_{J}.\label{eq:Liouvillian}\end{equation}
The solution of Eq. (\ref{eq:Liouville-v-N}) can be written in terms
of the evolution superoperator ${\cal U}(t)=\exp\{-\frac{i}{\hbar}{\cal L}t\}$
as $W(t)={\cal U}(t)W(0)$. In interaction picture with respect to
${\cal L}_{0}$, we define \begin{equation}
{\cal L}_{J}(t)={\cal U}_{0}^{\dagger}(t){\cal L}_{J}{\cal U}_{0}(t),\label{eq:int_picture_L}\end{equation}
 and\begin{equation}
W^{(I)}(t)={\cal U}_{0}^{\dagger}(t)W(0),\label{eq:int_picture_W}\end{equation}
where ${\cal U}_{0}(t)=\exp\{-\frac{i}{\hbar}{\cal L}_{0}t\}$. Using
the usual projection operator method ($P^{2}=P$, $Q=1-P$) we get\[
\frac{\partial}{\partial t}PW^{(I)}(t)=-iP{\cal L}_{J}(t)PW^{(I)}(t)-iP{\cal L}_{J}(t)QW(t_{0})\]
 \begin{equation}
-\int\limits _{t_{0}}^{t}d\tau P{\cal L}_{J}(t)Q{\cal L}_{J}(\tau)PW^{(I)}(\tau),\label{eq:QME}\end{equation}
where we have already truncated the perturbation expansion in ${\cal L}_{J}$
at the second order. Although Eq. (\ref{eq:QME}) is valid for an
arbitrary projector $P$, for the quality of the second order approximation,
the choice of projection operator $P$ is crucial. The best guidance
to this choice is provided by the physical conditions at which the
equation is applied. In optical spectroscopy, we often deal with systems
that are in the electronic ground state at the initial time, and the
bath DOF are relaxed into the canonical equilibrium. Thus, $W(t_{0})=|g\rangle\langle g|W_{eq}=\rho_{gg}W_{eq}$.
According to the Condon principle, the bath part of this initial condition
is not changed upon an ultrafast photo-excitation and the initial
condition for relaxation of the nuclear DOF is still given by a canonical
density matrix. As a result, the projection operator \begin{equation}
PA=Tr_{Q}\{A\}W_{eq},\label{eq:ProjOp}\end{equation}
has the convenient property $QW(t_{0})=0$, and the so-called initial
term $iP{\cal L}_{J}(t)QW(t_{0})$ is identically equal to zero. This
is equivalent to the statement that the projection by $P$ does not
lead to a loss of information about the system at $t_{0}$.

Another physically important situation occurs when the system has
already spent some time in the electronically excited state, and the
bath DOF at an excited electronic state $|a\rangle$ has relaxed into
local equilibrium represented by the density matrix $W_{eq}^{a}$.
The projection operator which eliminates the initial term is \begin{equation}
PA=\sum_{a}\langle a|Tr_{Q}\{A\}|a\rangle W_{eq}^{a}|a\rangle\langle a|.\label{eq:ProjOpF}\end{equation}
This discussion shows that, in general, it would be desirable that
the projection operator $P$ were time-dependent. It is indeed possible
to formulate corresponding time-dependent projection operator technique
rigorously (see e.g. Ref. \cite{Linden}). In this contribution, we
will not discuss such a time evolution of the projection operator,
and we will keep in mind that the present formulation is valid only
for not too long relaxation times. We will however discuss Eq. (\ref{eq:QME})
with projection operator, Eq. (\ref{eq:ProjOpF}), later in Section
\ref{sub:Long-time-limit} to demonstrate the relation between the
theory developed here and the FRET.

The main feature of the present theory is that the resonance coupling
Hamiltonian $H_{J}$ is the term with respect to which we apply perturbation
theory. If the perturbation theory is applied with respect to the
system--bath interaction Hamiltonian with projection operator given
by Eq. (\ref{eq:ProjOp}), we arrive at the Redfield type relaxation
equations (see e.g. \cite{MayKuehn}), while using the projector,
Eq. (\ref{eq:ProjOpF}), leads to so-called Modified Redfield equations
\cite{MukamelModRed,Yang02}.

\subsection{Projector vs. interaction picture\label{sub:Projector-vs.-interaction}}

Eq. (\ref{eq:QME}) is an equation of motion for the RDM. However,
the projection operator is applied to the interaction picture density
matrix $W^{(I)}(t)$ and not to $W(t)$ as one would expect. We therefore
need to express the evolution of $\rho(t)=Tr_{Q}\{W(t)\}$ in terms
of the evolution of $\bar{\rho}(t)=Tr_{Q}\{W^{(I)}(t)\}$. Because
Hamiltonian operators $H_{el}$ and $H_{B}$ act on different Hilbert
spaces, they commute and we can write\begin{equation}
\bar{\rho}(t)=U_{el}^{\dagger}(t)Tr_{Q}\{U_{B}^{\dagger}(t)W(t)U_{B}(t)\}U_{el}(t).\label{eq:RDM_int}\end{equation}
 Matrix elements $\bar{\rho}_{ab}(t)$ of the RDM therefore read\begin{equation}
\bar{\rho}_{ab}(t)=e^{i\omega_{ab}t}Tr_{Q}\{U_{a}^{\dagger}(t)W_{ab}(t)U_{b}(t)\}.\label{eq:RDM_int2}\end{equation}
 Due to the properties of the trace operation, we find that for populations\begin{equation}
\bar{\rho}_{aa}(t)=\rho_{aa}(t).\label{eq:RDM2RDM}\end{equation}
 For the coherences $\bar{\rho}_{ab}(t)$, $a\neq b$ it is not possible
to write directly such a simple result. However, an approximate relation
between the two RDMs can be established by studying the case where
$H_{J}=0$. In this case, $W_{ab}(t)=U_{B}(t)W_{eq}U_{B}^{\dagger}(t)$
and $\bar{\rho}_{ab}(t)=\bar{\rho}_{ab}(0)$, because all the time
evolution was accounted for by the interaction picture. At the same
time, however, we can show that in second cumulant approximation\begin{equation}
\rho_{ab}(t)=\rho_{ab}(0)e^{-i\omega_{ab}t-g_{a}(t)-g_{b}^{*}(t)},\; a\neq b,\label{eq:RDM_free1}\end{equation}
where $g_{a}(t)$ is the well-known line shape function associated
with the transition $|g_{a}\rangle\rightarrow|e_{a}\rangle$\cite{MukamelBook}.
Thus, because the purpose of the interaction picture is to suppress
the time evolution due to Hamiltonian $H_{0}$ we can write\begin{equation}
\bar{\rho}_{ab}(t)=e^{i\omega_{ab}t+(1-\delta_{ab})[g_{a}(t)+g_{b}^{*}(t)]}\rho_{ab}(t),\label{eq:int_in_RDM}\end{equation}
 so that the exponential prefactor compensates the {}``$J$-free''
evolution of the RDM elements in the spirit of the interaction picture.

\subsection{Reduced density matrix equations}

We can now rewrite Eq. (\ref{eq:QME}) in second order in terms of
$\bar{\rho}(t)$ as\[
\frac{\partial}{\partial t}\bar{\rho}(t)W_{eq}=-iTr_{Q}\{H_{J}(t)W_{eq}\}\bar{\rho}(t)W_{eq}\]
\begin{equation}
+i\bar{\rho}(t)Tr_{Q}\{W_{eq}H_{J}(t)\}W_{eq}+R(J^{2}),\label{eq:EM_1st_1}\end{equation}
 where \begin{equation}
R(J^{2})=\int\limits _{t_{0}}^{t}d\tau P{\cal L}_{J}(t)Q{\cal L}_{J}(\tau)PW^{(I)}(\tau).\label{eq:Second_ord}\end{equation}
 In this subsection, let us use the lower case indices to denote both
the one- and two-excitation states. The traces in Eq. (\ref{eq:EM_1st_1})
can be easily evaluated in second order cumulant approximation\[
\langle a|Tr_{Q}\{H_{J}(t)W_{eq}\}|b\rangle=J_{ab}e^{i\omega_{ab}t}Tr_{Q}\{U_{a}^{\dagger}(t)U_{b}(t)W_{eq}\}\]
 \begin{equation}
=J_{ab}e^{i\omega_{ab}t-(1-\delta_{ab})[g_{a}^{*}(t)+g_{b}(t)]}\equiv J_{ab}(t),\label{eq:Jt}\end{equation}
 and we arrive at \[
\frac{\partial}{\partial t}\bar{\rho}_{ab}(t)=-\frac{i}{\hbar}\sum_{c}J_{ac}(t)\bar{\rho}_{cb}(t)\]
 \begin{equation}
+\frac{i}{\hbar}\sum_{c}\bar{\rho}_{ac}(t)J_{cb}(t)-R(J^{2})/W_{eq}.\label{eq:EM_1st_A}\end{equation}
In the second order term, Eq. (\ref{eq:Second_ord}), we have to evaluate
two commutators of Hamiltonian $H_{J}$ with the reduced density matrix\[
R(J^{2})/W_{eq}=\frac{1}{\hbar^{2}}\int\limits _{t_{0}}^{t}d\tau\Big[Tr_{Q}\{H_{J}(t)H_{J}(\tau)W_{eq}\}\bar{\rho}(\tau)\]
 \[
-Tr_{Q}\{H_{J}(t)W_{eq}\}Tr_{Q}\{H_{J}(\tau)W_{eq}\}\bar{\rho}(\tau)\]
\[
-Tr_{Q}\{H_{J}(t)\bar{\rho}(\tau)W_{eq}H_{J}(\tau)\}\]
 \[
+Tr_{Q}\{H_{J}(t)W_{eq}\}\bar{\rho}(\tau)Tr_{Q}\{W_{eq}H_{J}(\tau)\}\]
\[
-Tr_{Q}\{H_{J}(\tau)\bar{\rho}(\tau)W_{eq}H_{J}(t)\}\]
 \[
+Tr_{Q}\{H_{J}(\tau)W_{eq}\}\bar{\rho}(\tau)Tr_{Q}\{W_{eq}H_{J}(t)\}\]
\[
+\bar{\rho}(\tau)Tr_{Q}\{W_{eq}H_{J}(\tau)H_{J}(t)\}\]
 \begin{equation}
-\bar{\rho}(\tau)Tr_{Q}\{W_{eq}H_{J}(\tau)\}Tr_{Q}\{W_{eq}H_{J}(t)\}\Big].\label{eq:R2_gen}\end{equation}
 The matrix elements involved in Eq. (\ref{eq:R2_gen}) read in detail\[
\langle a|Tr_{Q}\{H_{J}(t)H_{J}(\tau)W_{eq}\}|b\rangle=\sum_{c}J_{ac}J_{cb}e^{i\omega_{ac}t+i\omega_{cb}\tau}\]
\begin{equation}
\times Tr_{Q}\{U_{a}^{\dagger}(t)U_{c}(t)U_{c}^{\dagger}(\tau)U_{b}(\tau)W_{eq}\},\label{eq:J2_1}\end{equation}
\[
\langle a|Tr_{Q}\{H_{J}(t)|c\rangle\dots W_{eq}\langle d|H_{J}(\tau)\}|b\rangle=J_{ac}\dots J_{db}\]
 \begin{equation}
\times e^{i\omega_{ac}t+i\omega_{dc}\tau}Tr_{Q}\{U_{a}^{\dagger}(t)U_{c}(t)W_{eq}U_{d}^{\dagger}(\tau)U_{b}(\tau)\}\label{eq:J2_2}\end{equation}
and\[
\langle a|Tr_{Q}\{W_{eq}H_{J}(\tau)H_{J}(t)\}|b\rangle=\sum_{c}J_{ac}J_{cb}e^{i\omega_{ac}\tau+i\omega_{cb}t}\]
 \begin{equation}
\times Tr_{Q}\{U_{a}^{\dagger}(\tau)U_{c}(\tau)U_{c}^{\dagger}(t)U_{b}(t)W_{eq}\}.\label{eq:J2_3}\end{equation}
 We introduce an auxiliary function\[
M_{abcd}(t,\tau)=Tr_{Q}\{U_{a}^{\dagger}(t)U_{b}(t)\]
\begin{equation}
\times U_{c}^{\dagger}(\tau)U_{d}(\tau)W_{eq}\}e^{i\omega_{ab}t+i\omega_{cd}\tau},\label{eq:Mfce}\end{equation}
 with the property\begin{equation}
M_{abcd}(t,\tau)=M_{dcba}^{*}(\tau,t)\label{eq:Mstart}\end{equation}
 and after changing the integration variable to $\tau'=t-\tau$ we
can write the EM in the form\[
\frac{\partial}{\partial t}\bar{\rho}_{ab}(t)=-\frac{i}{\hbar}\sum_{c}J_{ac}(t)\bar{\rho}_{cb}(t)\]
 \[
+\frac{i}{\hbar}\sum_{c}\bar{\rho}_{ac}(t)J_{cb}(t)-\sum_{cd}\frac{1}{\hbar^{2}}\int\limits _{0}^{t-t_{0}}d\tau\]
 \[
\left[J_{ac}J_{cd}M_{accd}(t,t-\tau)-J_{ac}(t)J_{cd}(t-\tau)\right]\bar{\rho}_{db}(t-\tau)\]
 \[
-\left[J_{ca}^{*}J_{bd}^{*}M_{cabd}^{*}(t,t-\tau)-J_{ca}^{*}(t)J_{bd}^{*}(t-\tau)\right]\bar{\rho}_{cd}(t-\tau)\]
 \[
-\left[J_{db}J_{ac}M_{dbac}(t,t-\tau)-J_{db}(t)J_{ac}(t-\tau)\right]\bar{\rho}_{cd}(t-\tau)\]
\[
+[J_{bd}^{*}J_{dc}^{*}M_{bddc}^{*}(t,t-\tau)\]
 \begin{equation}
-J_{bd}^{*}(t)J_{dc}^{*}(t-\tau)]\bar{\rho}_{ac}(t-\tau).\label{eq:RJ2_nonmarkov_final}\end{equation}
This is a form suitable for introducing both long time limit $(t_{0}\rightarrow\infty)$
and the Markov approximation $(\bar{\rho}_{ab}(t-\tau)\approx\bar{\rho}_{ab}(t)$).
We apply only the later one, because we are interested mostly in short
times. Now, it only remains to evaluate the four-index matrix $M_{abcd}(t,\tau)$
which is done in Appendix A using the cumulant expansion. In Markov
approximation, we can write our equations as\[
\frac{\partial}{\partial t}\bar{\rho}_{ab}(t)=-\frac{i}{\hbar}\sum_{c}J_{ac}(t)\bar{\rho}_{cb}(t)\]
 \[
+\frac{i}{\hbar}\sum_{c}\bar{\rho}_{ac}(t)J_{cb}(t)-\sum_{cd}\Big[R_{accd}(t)\bar{\rho}_{db}(t)-R_{cabd}^{*}(t)\bar{\rho}_{cd}(t)\]
 \begin{equation}
-R_{dbac}(t)\bar{\rho}_{cd}(t)+R_{bddc}^{*}(t)\bar{\rho}_{ac}(t)\Big],\label{eq:RJ2_markov}\end{equation}
 where we introduced relaxation tensor\[
R_{abcd}(t)=\frac{1}{\hbar^{2}}\int\limits _{0}^{t}d\tau\Big[J_{ab}J_{cd}M_{abcd}(t,t-\tau)\]
\begin{equation}
-J_{ab}(t)J_{cd}(t-\tau)\Big].\label{eq:RelMatr}\end{equation}

We derived EM of the RDM in the interaction picture. We use Eq. (\ref{eq:int_in_RDM})
to transform the RDM in Schr\"{o}dinger picture if necessary. The
first two terms in Eq. (\ref{eq:RJ2_markov}) correspond to the delocalization
effect of resonance coupling given by Eq. (\ref{eq:Jt}). While the
term $e^{i\omega_{ab}t}$ in Eq. (\ref{eq:Jt}) originates from the
interaction picture with respect to the electronic Hamiltonian, the
presence of the line-shape functions in Eq. (\ref{eq:Jt}) shows that
the magnitude of this coupling decreases exponentially with growing
time $t$. Thus the bath fluctuations dynamically destroys the resonance
coupling.

\subsection{Two-excitation states}

In higher order spectroscopies, ESA contributes considerably to the
signal. A delicate balance of ground state bleaching and stimulated
emission on one hand, and the ESA on the other hand, is behind the
disappearance of the 2D crosspeaks when resonance coupling goes to
zero. For weakly coupled aggregates, correct description of ESA, and
correspondingly the dephasing of coherences between one-excitation
and two-excitation state is indispensable. 

In this subsection, let us again denote one-excitation and two-excitation
states by the lower case and the upper case letters, respectively.
The evolution of the system in a two-exciton state $A=(a,b)$ is described
by the evolution operator\[
U_{A}(t)=\exp\left\{ -\frac{i}{\hbar}H_{A}t\right\} \]
 \begin{equation}
=\exp\left\{ -\frac{i}{\hbar}[H_{a}\otimes1_{\{a\}}+1_{\{b\}}\otimes H_{b}]t\right\} ,\label{eq:UA}\end{equation}
 i.e.\begin{equation}
U_{A}(t)=U_{a}(t)\otimes U_{b}(t)\otimes1_{\{a,b\}}.\label{eq:UAasUaUb}\end{equation}
 We denoted the direct product of unity operators from Hilbert spaces
except of those in a set $\{a,b,\dots\}$ by $1_{\{a,b,\dots\}}$.
Expressions containing a product $U_{A}(t)U_{B}^{\dagger}(t)$, where
one of the excited states is shared by the two-exciton states $A$
and $B=(a,c)$, thus yield\begin{equation}
U_{(a,b)}(t)U_{(a,c)}^{\dagger}(t)=1_{\{b,c\}}\otimes U_{b}(t)U_{c}^{\dagger}(t).\label{eq:two_ex_evol}\end{equation}
 Considering the first order term and a general case $M=(m,n)$, $N=(k,l)$
one arrives at\[
J_{MN}(t)=J_{MN}e^{i\omega_{MN}t}Tr_{Q}\{U_{M}^{\dagger}(t)U_{N}(t)W_{eq}\}\]
 \begin{equation}
=\delta_{mk}J_{nl}(t)+\delta_{nl}J_{mk}(t)+\delta_{ml}J_{nk}(t)+\delta_{nk}J_{ml}(t).\label{eq:JMNt}\end{equation}
 For the terms in the second order of $J$ we have in a complete analogy\[
J_{MN}J_{cd}M_{MNcd}(t,\tau)=\]
 \[
\delta_{mk}J_{nl}J_{cd}M_{nlcd}(t,\tau)+\delta_{nl}J_{mk}J_{cd}M_{mkcd}(t,\tau)\]
 \begin{equation}
+\delta_{ml}J_{nk}J_{cd}M_{nkcd}(t,\tau)+\delta_{nk}J_{ml}J_{cd}M_{mlcd}(t,\tau),\label{eq:MMNcd}\end{equation}
 and consequently\[
R_{MNcd}(t)=\delta_{mk}R_{nlcd}(t)+\delta_{nl}R_{mkcd}(t)\]
\begin{equation}
+\delta_{ml}R_{nkcd}(t)+\delta_{nk}R_{mlcd}(t).\label{eq:RMNcd}\end{equation}
 Thus, all quantities corresponding to the two-excitation states can
be expressed directly using the one-excitation quantities.

\subsection{Homogeneous limit\label{sub:Homogenous-limit}}

To evaluate the relaxation tensor we need to evaluate the following
two expressions\begin{equation}
R_{abcd}^{\prime}(t)=\frac{1}{\hbar^{2}}J_{ab}J_{cd}\int\limits _{0}^{t}d\tau M_{abcd}(t,t-\tau),\label{eq:Rprime}\end{equation}
 and\begin{equation}
R_{abcd}^{\prime\prime}(t)=-\frac{1}{\hbar^{2}}J_{ab}(t)\int\limits _{0}^{t}d\tau J_{cd}(t-\tau).\label{eq:Rprime2}\end{equation}
 In order to simplify the equations, we will assume so-called homogeneous
limit, where we have \begin{equation}
g_{a}(t)=\Gamma_{a}t.\label{eq:ga_hom}\end{equation}
 This simple formula allows us to evaluate all terms in the relaxation
matrix analytically. First we observe that\begin{equation}
J_{ab}(t)=J_{ab}e^{-(\Gamma_{a}+\Gamma_{b})t+i\omega_{ab}t},\label{eq:Jhom}\end{equation}
 and therefore\[
R_{abcd}^{\prime\prime}(t)=J_{ab}J_{cd}e^{-(\Gamma_{a}+\Gamma_{b})t+i\omega_{ab}t}\frac{1}{(\Gamma_{c}+\Gamma_{d})-i\omega_{cd}}\]
\begin{equation}
\times\left(e^{-(\Gamma_{c}+\Gamma_{d})t+i\omega_{cd}t}-1\right).\label{eq:R2prim_hom}\end{equation}
 The $R^{\prime}$ elements are obtained in a similar manner. We start
with a splitting of the $M$ functions\[
M_{abcd}(t,\tau)=M_{abcd}^{\prime}(t)M_{abcd}^{\prime\prime}(t-\tau)\]
 \begin{equation}
\times M_{abcd}^{\prime\prime\prime}(\tau)e^{i\omega_{ab}t+i\omega_{cd}\tau}.\label{eq:Meq3M}\end{equation}
Such a splitting is possible for an arbitrary $g(t)$ function and
is not limited to the homogeneous limit. In general, the integrals,
Eqs. (\ref{eq:Rprime}) and (\ref{eq:Rprime2}), can be evaluated
using the Fourier transform. In homogeneous limit we find that\[
M_{abcd}^{\prime}(t)=e^{-\alpha_{abcd}t},\; M_{abcd}^{\prime\prime}(t)=e^{-\beta_{abcd}t},\;\]

\begin{equation}
M_{abcd}^{\prime\prime\prime}(t)=e^{-\gamma_{abcd}t},\label{eq:3M}\end{equation}
 where\begin{equation}
\alpha_{abcd}=(1-\delta_{ac}+\delta_{ad})\Gamma_{a}+(1-\delta_{bc}-\delta_{bd})\Gamma_{b},\label{eq:alpha}\end{equation}
 \begin{equation}
\beta_{abcd}=(\delta_{ad}-\delta_{ac})\Gamma_{a}+(\delta_{bc}-\delta_{bd})\Gamma_{b},\label{eq:beta}\end{equation}
 and\begin{equation}
\gamma_{abcd}=\Gamma_{c}+\Gamma_{d}+(\delta_{ac}-\delta_{ad})\Gamma_{a}+(\delta_{bd}-\delta_{bc})\Gamma_{b}.\label{eq:gamma}\end{equation}
Using the definition, Eq. (\ref{eq:Rprime}), we get\[
R_{abcd}^{\prime}(t)=J_{ab}J_{cd}e^{i(\omega_{ab}+\omega_{cd})t-(\alpha_{abcd}+\gamma_{abcd})t}\]
\begin{equation}
\times\frac{1}{\gamma_{abcd}-\beta_{abcd}-i\omega_{cd}}\left[e^{(\gamma_{abcd}-\beta_{abcd})t-i\omega_{cd}t}-1\right].\label{eq:Rprime_hom}\end{equation}
The case when $\gamma_{abcd}-\beta_{abcd}-i\omega_{cd}=0$, which
can occur for homo aggregates, has to be considered separately. According
to Eqs. (\ref{eq:Rprime}) and (\ref{eq:Meq3M}), when the denominator
is equal to zero, the dependence on the integration variable disappears
and the integral leads to $t$. Consequently,\begin{equation}
R_{abcd}^{\prime}(t)=J_{ab}J_{cd}te^{i(\omega_{ab}+\omega_{cd})t-(\alpha_{abcd}+\gamma_{abcd})t}.\label{eq:Rprime_degen}\end{equation}
If the dephasing constants $\Gamma$ are non-zero, no such problem
can occur with Eq. (\ref{eq:R2prim_hom}).

We stress that the homogeneous limit is used here for demonstration
purposes only. The line broadening function, Eq. (\ref{eq:ga_hom}),
corresponds to a limit of ultrafast stochastic bath with correlation
function $C(t)=\Gamma\delta(t)$. Such correlation function does not
allow for introducing required thermodynamic properties $C(\omega)=e^{\hbar\omega/k_{B}T}C(-\omega)$
for the corresponding spectral density $C(\omega)=\int_{-\infty}^{\infty}dtC(t)e^{i\omega t}$
. Consequently, the relaxation tensor given by Eqs. (\ref{eq:R2prim_hom})
and (\ref{eq:Rprime_hom}) does not lead to any finite temperature
thermal equilibrium. We will treat calculations with realistic correlation
functions elsewhere.

\section{Long-time limit of Equations of Motion\label{sub:Long-time-limit}}

In order to establish the relation between our EM, Eq. (\ref{eq:RJ2_markov}),
and standard EM used to describe dissipative dynamics and energy transfer,
we show that the above derivation leads to the well-know Förster resonance
transfer rates in the long time limit. Considering the projection
operator, Eq. (\ref{eq:ProjOpF}), we first find that \begin{equation}
P{\cal L}_{J}(t)PW^{(I)}(t)=0,\label{eq:PLPnull}\end{equation}
 which leads also to\[
P{\cal L}_{J}(t)Q{\cal L}_{J}(\tau)PW^{(I)}(\tau)\]
 \begin{equation}
=P{\cal L}_{J}(t){\cal L}_{J}(\tau)PW^{(I)}(\tau).\label{eq:kernelPF}\end{equation}
 Using the definition of the projection operator, Eq. (\ref{eq:ProjOpF}),
Eq. (\ref{eq:QME}) turns into\[
\frac{\partial}{\partial t}\rho_{aa}(t)=-\frac{1}{\hbar^{2}}\sum_{b}|J_{ab}|^{2}\]
 \[
\times\int\limits _{t_{0}}^{t}d\tau\Big[\{C_{ba}^{*}(t-\tau)+C_{ba}(t-\tau)\}\rho_{aa}(\tau)\]
 \begin{equation}
-\{C_{ab}(t-\tau)+C_{ab}^{*}(t-\tau)\}\rho_{bb}(\tau)\Big],\label{eq:popEq}\end{equation}
 where \begin{equation}
C_{ab}(t)=Tr_{Q}\left\{ U_{a}(t)U_{b}^{\dagger}(t)W_{eq}^{ab}\right\} .\label{eq:Cab}\end{equation}
The initial state of the bath $W_{eq}^{ab}$ is the one in which chromophore
$b$ is vibrationally relaxed in the electronically excited state,
and $a$ is relaxed the electronic groundstate. Thus, $W_{eq}^{ab}=W_{eq}^{g}W_{eq}^{b}$.
We apply the second order cumulant expansion and get\[
C_{ab}(t)=Tr_{Q}\left\{ U_{a}(t)U_{g}^{\dagger}(t)W_{eq}^{q}\right\} Tr_{Q}\left\{ U_{g}(t)U_{b}^{\dagger}(t)W_{eq}^{b}\right\} \]
 \begin{equation}
=e^{-g_{a}(t)-i\omega_{ag}t}e^{-g_{b}^{*}(t)+i(\omega_{bg}-2\lambda_{b})t},\label{eq:Cab_inG}\end{equation}
 where we used the property $2\lambda_{a}=Tr_{Q}\{\Delta V_{a}W_{g}^{eq}\}-Tr_{Q}\{\Delta V_{a}W_{a}^{eq}\}$
(see e.g. Ref. \cite{MukamelBook}). By substitution $\tau'=t-\tau$
, limit $t_{0}\rightarrow-\infty$ and Markov approximation $\rho_{aa}(t-\tau^{\prime})\approx\rho_{aa}(t)$
we obtain relaxation rate equation with Förster rates (see e.g. Ref.
\cite{Yang02})\[
K_{a\leftarrow b}=2Re\left|\frac{J_{ab}}{\hbar}\right|^{2}\]
\begin{equation}
\times\int\limits _{0}^{\infty}d\tau e^{-g_{a}(\tau)-i\omega_{ag}t-g_{b}^{*}(\tau)+i(\omega_{bg}-2\lambda_{b})\tau}.\label{eq:Foerster}\end{equation}
This demonstrates the relation of Eq. (\ref{eq:RJ2_markov}) to the
Förster energy transfer rates. It is important to point out that the
only difference between the two sets of equations is in the {}``initial''
condition set on the bath part of the density matrix by the choice
of the projection superoperator. Thanks to our choice of the projection
operator, equations used in this work retain the description of the
electronic coherences, while F\"{o}rster rate equations give no prescription
for them.

\section{Absorption and Two-dimensional Correlation Photon Echo Spectra\label{sec:Two-dimensional-Correlation-Photon}}

To calculate optical spectra we concentrate on coherence elements
of the RDM that correspond to optical transitions. We apply perturbation
theory with respect to electric field of an incident light to calculate
absorption and two-dimensional correlation photon echo spectra \cite{MukamelBook}.

\subsection{Optical coherences\label{sub:Optical-coherences}}

To calculate response functions needed for evaluations of optical
spectra in general, we need first to calculate evolution operators
${\cal U}(t)$ which fulfill the relation\begin{equation}
\rho(t)={\cal U}(t)\rho(0).\label{eq:evolop}\end{equation}
We use a simple consequence of this equation, namely\begin{equation}
{\cal U}_{abcd}(t)=\rho_{ab}(t),\label{eq:practical_U_def}\end{equation}
 where $\rho_{ab}(t)$ is calculated using EM, Eq. (\ref{eq:RJ2_markov}),
with initial condition $\rho_{ab}(0)=\delta_{ac}\delta_{bd}.$ 

If any of the indices $a,b,c,d$ equals $g$, we have $R_{abcd}=0$.
For the optical coherences involving the ground state we therefore
obtain the following EM\[
\frac{\partial}{\partial t}\bar{\rho}_{ag}(t)=-\sum_{c}\frac{i}{\hbar}J_{ac}(t)\]
\begin{equation}
+\sum_{c}\sum_{d\neq a,c}R_{addc}(t)\bar{\rho}_{cg}(t).\label{eq:EM_cohg}\end{equation}
 Also, when both the single and double excited states are present
in the $R_{abcd}$ matrix, many elements of the matrix are zero, most
importantly those where e.g. $a$ and $b$ are a single excitation
and double excitation indices, respectively. For the coherences involving
the one exciton and two exciton states we have EM\[
\frac{\partial}{\partial t}\bar{\rho}_{aB}(t)=-\frac{i}{\hbar}\sum_{c}J_{ac}(t)\bar{\rho}_{cB}(t)+\frac{i}{\hbar}\sum_{C}\bar{\rho}_{aC}(t)J_{CB}(t)\]
\[
-\sum_{cd}R_{accd}(t)\bar{\rho}_{dB}(t)+\sum_{cD}\left[R_{caBD}^{*}(t)+R_{DBac}(t)\right]\bar{\rho}_{cD}(t)\]
 \begin{equation}
-\sum_{CD}R_{BDDC}^{*}(t)\bar{\rho}_{aC}(t).\label{eq:EM_coh2e}\end{equation}
 It is possible to rewrite this equation entirely using the one-excitation
indices by considering Eqs. (\ref{eq:JMNt}) and (\ref{eq:RMNcd}).
Setting $B=(\sigma,\pi)$, $D=(\gamma,\delta)$ and $C=(\alpha,\beta)$
yields \[
\frac{\partial}{\partial t}\bar{\rho}_{a(\sigma\pi)}(t)=-\frac{i}{\hbar}\sum_{c}J_{ac}(t)\bar{\rho}_{c(\sigma\pi)}(t)\]
 \begin{equation}
-\sum_{cd}R_{accd}(t)\bar{\rho}_{d(\sigma\pi)}(t)+{\cal T}_{1}+{\cal T}_{2}-{\cal T}_{3},\label{eq:EM_coh2e_1x}\end{equation}
where last three terms ${\cal T}_{1}$, ${\cal T}_{2}$ and ${\cal T}_{3}$
are somewhat lengthy. We present the first term here,\[
{\cal T}_{1}=\frac{i}{\hbar}\Big[\sum_{\beta=\sigma+1}^{K}\bar{\rho}_{a(\sigma\beta)}(t)J_{\beta\pi}(t)+\sum_{\alpha=1}^{\pi-1}\bar{\rho}_{a(\alpha\pi)}(t)J_{\alpha\sigma}(t)\]
\begin{equation}
+\sum_{\beta=\pi+1}^{K}\bar{\rho}_{a(\pi\beta)}(t)J_{\beta\sigma}(t)+\sum_{\alpha=1}^{\sigma-1}\bar{\rho}_{a(\alpha\sigma)}(t)J_{\alpha\pi}(t)\Big],\label{eq:T1}\end{equation}
and the remaining two are presented in full in the Appendix. From
the point of view of simulation feasibility, Eq. (\ref{eq:EM_coh2e_1x})
represents the main advantage of treating a weakly coupled excitonic
systems in the site basis, as opposed to the treatment in the excitonic
basis. Although Eq. (\ref{eq:EM_coh2e_1x}) is rather lengthy, one
is concerned only with tensor quantities with the number of elements
proportional to $\sim N^{4}$, where $N$ is the number of chromophores,
as opposed to $\sim N^{6}$ which would be required in excitonic basis.

\subsection{Absorption spectrum}

The absorption spectrum is given by expression\[
\alpha(\omega)\approx\frac{\omega}{n(\omega)}Re\int\limits _{0}^{\infty}dte^{i\omega t}\]
\begin{equation}
\times\left\langle \sum_{ab}d_{ga}{\cal U}_{agbg}(t)d_{bg}\rho_{gg}\right\rangle ,\label{eq:abs}\end{equation}
 where $\langle\dots\rangle$ represents an averaging over isotropic
distribution of orientations of the molecular transitions with respect
to the light polarization. The transition dipole moments $d_{ag}$
have to be understood as projections of the transition dipole moments
on the light polarization vector ${\bf e}$, i.e. $d_{ag}={\bf d}_{ag}\cdot{\bf e}$.
The averaging is done over a product of two of such quantities. We
have \begin{equation}
\Omega_{ab}\equiv\langle({\bf d}_{ag}\cdot{\bf e})({\bf d}_{bg}\cdot{\bf e})\rangle_{orient.}=\frac{1}{3}\frac{{\bf d}_{ag}\cdot{\bf d}_{bg}}{|{\bf d}_{ag}||{\bf d}_{bg}|}.\label{eq:orient_abs}\end{equation}
 If one now defines $\bar{d}_{ag}$ to $\bar{d}_{ag}\equiv|{\bf d}_{ag}|$
one can write\[
\alpha(\omega)\approx\frac{\omega}{n(\omega)}Re\int\limits _{0}^{\infty}dte^{i\omega t}\]

\begin{equation}
\times\sum_{ab}\Omega_{ab}\bar{d}_{ga}\bar{d}_{bg}{\cal U}_{agbg}(t)\rho_{gg}.\label{eq:abs_Ome}\end{equation}
We use Eq. \ref{eq:abs_Ome} in subsequent simulations of absorption
spectra. It is important to note that because we do not work with
electronic eigenstates one cannot assume the so-called secular approximation
(${\cal U}_{abcd}(t)=\delta_{ac}\delta_{bd}{\cal U}_{abab}(t)$) to
be valid, and the orientational factor does not reduce to simple $1/3$.

\subsection{Two-dimensional spectrum\label{sub:Two-dimensional-spectrum}}

Two-dimensional (2D) Fourier transformed photon echo (FTPE) spectroscopy
is well described by the third order time dependent perturbation theory
with respect to light-matter interaction \cite{MukamelBook}. The
spectroscopic signals are expressed in terms of response functions
corresponding to light-matter interaction events. These response function
allow us to calculate an arbitrary third order response of a multi-level
electronic system, provided we know the evolution superoperators ${\cal U}$
and the transition dipole moment elements $d_{ij}$, for all involved
electronic levels. The response functions are directly proportional
to the observed signal if the incident pulses are infinitely short.
Expressions for all response functions $R_{ig}$ and $R_{if}$ ($i=1,\dots4$)
involved in the calculations of the 2D FTPE of our model systems are
summarized in Appendix \ref{sec:Appendix-B:-Third}.

In this paper we will consider only 2D spectra with zero population
time, $t_{2}=0$. The impulsive limit signals in the rephasing and
non-rephasing configuration can be obtained as follows (see e.g. \cite{Brixner04b})\[
S_{R}(t_{3},t_{1})=R_{2g}(t_{3},0,t_{1})\]
\begin{equation}
+R_{3g}(t_{3},0,t_{1})-R_{1f}^{*}(t_{3},0,t_{1}),\label{eq:SR}\end{equation}
 and\[
S_{NR}(t_{3},t_{1})=R_{1g}(t_{3},0,t_{1})\]
\begin{equation}
+R_{4g}(t_{3},0,t_{1})-R_{2f}(t_{3},0,t_{1}).\label{eq:SNR}\end{equation}
The total signal is given by\begin{equation}
S(t_{3},t_{1})=\Theta(t_{1})S_{R}(t_{3},t_{1})+\Theta(-t_{1})S_{NR}(t_{3},-t_{1}),\label{eq:totSig}\end{equation}
 and consequently, the 2D spectrum, which is defined as a double Fourier
transform of the signal $S(t_{3},t_{1})$ is given by \cite{Brixner04b}\[
\Xi(\omega_{3},\omega_{1})=\int\limits _{0}^{\infty}dt_{3}\int\limits _{0}^{\infty}dt_{1}S_{R}(t_{3},t_{1})e^{i\omega_{3}t_{3}-i\omega_{1}t_{1}}\]
 \begin{equation}
+\int\limits _{0}^{\infty}dt_{3}\int\limits _{0}^{\infty}dt_{1}S_{NR}(t_{3},t_{1})e^{i\omega_{3}t_{3}+i\omega_{1}t_{1}}.\label{eq:2Dspec}\end{equation}
 In each response function component of the 2D spectrum, Fourier transform
can be performed in $t_{1}$ and $t_{3}$ times separately.

\section{Results and Discussion\label{sec:Results-and-Discussion}}

\begin{figure}
\includegraphics[width=1\columnwidth]{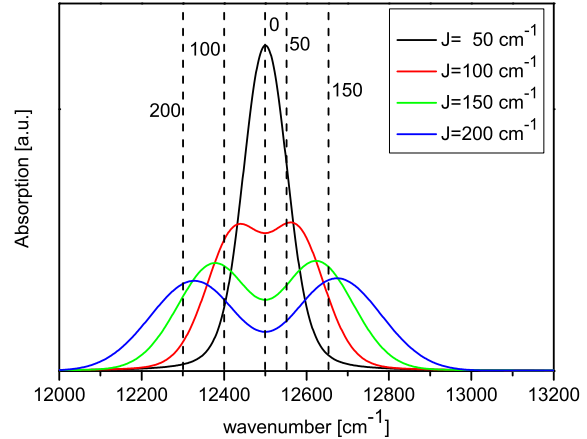}

\caption{\label{fig:Comparison-of-linear}Illustration of excitonic splitting
in a model homodimer system. Linear absorption spectrum of a homodimer
with perpendicular transition dipole moments calculated using Eq.
(\ref{eq:abs_Ome}). The following parameters were used to illustrate
the influence of resonance coupling: $\epsilon_{1}=\epsilon_{2}=12500$
cm$^{-1}$, resonance coupling $J=50$, $100$, $150$ and $200$
cm$^{-1}$, and $\Gamma=1/400$ fs$^{-1}$. }

\end{figure}
The theory developed in the above sections has been implemented in
the spectroscopic package NOSE \cite{NOSE} which was used to perform
simulations of impulsive limit 2D spectra of several small model systems.
We have chosen systems where effects of weak excitonic coupling, such
as those reported in Ref. \cite{Zigmantas06a} for LH3 could be expected.
We demonstrate below that our local basis description is sufficient
to account for such effects. To take advantage of analytic equations
derived in earlier sections, we stay in homogeneous limit. Simulations
taking advantage of the full description, including a finite bath
correlation time will be presented elsewhere.

\subsection{Molecular dimer\label{sub:Symmetric-homodimer}}

\begin{figure}
\includegraphics[width=1\columnwidth]{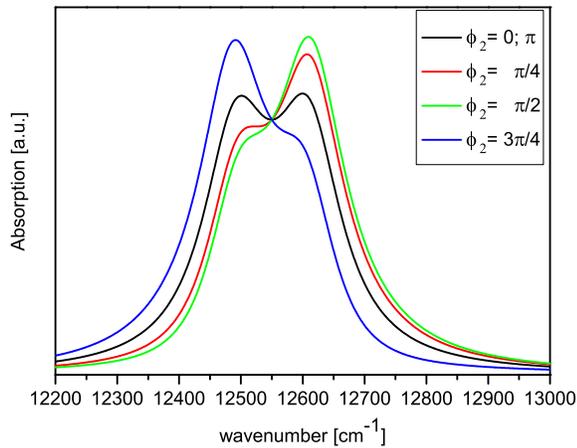}

\caption{\label{fig:redistribution}Illustration of oscillator strength redistribution
due to resonance coupling in a model heterodimer. The mutual position
and orientation of the transition dipoles are described by angles
$\phi_{1}$ and $\phi_{2}$ from Fig. \ref{fig:illustration}. The
parameters of the model are $\epsilon_{1}=12500$ cm$^{-1}$, $\epsilon_{2}=12600$
cm$^{-1}$, $J=50$ cm$^{-1}$, $\Gamma=1/100$ fs$^{-1}$, $\phi_{1}=\pi/2$
and $\phi_{2}=0$, $\pi/4$, $\pi/2$, $3\pi/2$ and $\pi$.}

\end{figure}

The simplest system where a weak excitonic coupling effect in excited
state absorption can be observed is a molecular dimer. Resonance interaction
leads to the splitting of the excited states, redistribution of the
transition dipole moments and a shift of excited state absorption.
These effects will be demonstrated here. In addition, one can also
expect energy transfer between the two split excitonic levels, formation
of a coherence between excitonic levels upon excitation by light and
its dephasing. This class of effects is associated with the evolution
of the system in the excited state band, and will be studied within
our model elsewhere. Fig. \ref{fig:illustration} illustrates the
dimer geometry and its excitonic splitting. In all dimers considered
here, transition dipole moments lie in a $z=0$ plane. Orientation
of the dipoles with respect to $x$ and $y$ axes is determined by
an angle $\phi$ such that $d_{x}=|d|\cos\phi$. 

Absorption spectrum of the dimer displays the splitting of the levels,
as well as the transition dipole moment redistribution. Fig. \ref{fig:Comparison-of-linear}
presents absorption spectra of a model homo-dimer with resonance coupling
varying from $50$ cm$^{-1}$ to $200$ cm$^{-1}$ and dephasing parameters
$\Gamma=1/400$ fs$^{-1}$. Because the monomeric transition energies
of the two levels are the same, excitonic mixing of the two levels
is maximal at any resonance coupling value. We have artificially chosen
the dipole moments perpendicular to each other to eliminate the effect
of transition dipole moment redistribution. We can see from Fig. \ref{fig:Comparison-of-linear}
that the prediction of the absorption maxima agrees rather well with
the prediction of excitonic model (splitting of $2J$). It can also
be noticed that the splitting is smaller than predicted by excitonic
theory when resonance coupling is small, most likely due to the bath
suppressing the exciton coupling term in Eq. (\ref{eq:Jt}).

The effect of the transition dipole moment redistribution is illustrated
on Fig. \ref{fig:redistribution}. A hetero-dimer with difference
of $100$ cm$^{-1}$ between the transition energies on the two monomers
was chosen, and the absorption spectrum was calculated for a fixed
resonance coupling value of $50$ cm$^{-1}$. Different mutual orientations
of the dipole moments lead to enhancement of the absorption on one
or the other split  level, depending on mutual orientation of the
molecules. 

\begin{figure}
\includegraphics[clip,width=1\columnwidth]{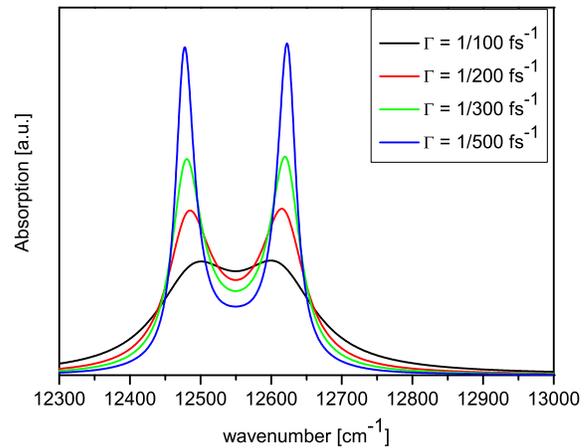}

\caption{\label{fig:Influence-of-dephasing} Influence of dephasing on excitonic
splitting in a model heterodimer. The parameters of the model are
$\epsilon_{1}=12600$ cm$^{-1}$, $\epsilon_{2}=12500$ cm$^{-1}$,
$J=50$ cm$^{-1}$, $\phi_{1}=\pi/2$ and $\phi_{2}=\pi$. The dephasing
rates are $\Gamma=1/100$, $1/200$, $1/300$ and $1/500$ fs$^{-1}$. }

\end{figure}

In Fig. \ref{fig:Influence-of-dephasing} we demonstrate that increasing
the dephasing rate $\Gamma$ leads to broadening of the absorption
spectrum. Unlike in case of exciton splitting were the position of
the line does not shift, here we can observe a small shift towards
less pronounced splitting with increasing the dephasing rate. 

\begin{figure}
\includegraphics[clip,width=1\columnwidth]{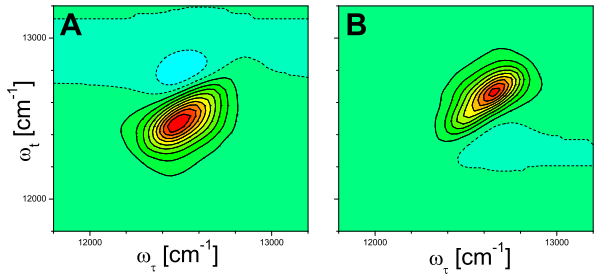}

\caption{\label{fig:Dimer-2d}2D spectrum of a homo dimer. Subfigures: (A)
in line configuration with excitonic coupling $J=-80$ cm$^{-1}$,
disorder width $\Delta=100$ cm$^{-1}$; (B) sandwich configuration
with excitonic coupling $J=40$ cm$^{-1}$ and $\Delta=100$ cm$^{-1}$. Full line contours represent positive values from 10\% to 100\% of the maximum. Zero and negative contours are dashed, and spaced by 10\% of the positive maximum. }

\end{figure}

The effect of excited state absorption offset cannot be demonstrated
on an ordinary absorption spectrum. We have therefore calculated 2D
electronic spectra at population time $t_{2}=0$ for two dimer configurations.
An \emph{in line} configuration, Fig. \ref{fig:Dimer-2d}A, corresponds
to two transition dipole moments oriented head-to-tail with the distance
and dipole moment length chosen such that the dipole-dipole coupling
leads to $J=-80$ cm$^{-1}$. This results in an offset of the ESA
towards higher frequencies. The \emph{sandwich} configuration with
two parallel dipole moments and the same center to center distance
results in positive coupling $J=40$ cm$^{-1}$ . The ESA appears
on the lower frequencies (Fig. \ref{fig:Dimer-2d}B) in this case.
Both calculations are performed with a diagonal Gaussian disorder
with the FWHM of $\Delta=100$ cm$^{-1}$.

\subsection{Small aggregates\label{sub:Trimer}}

\begin{figure}
\includegraphics[clip,width=1\columnwidth]{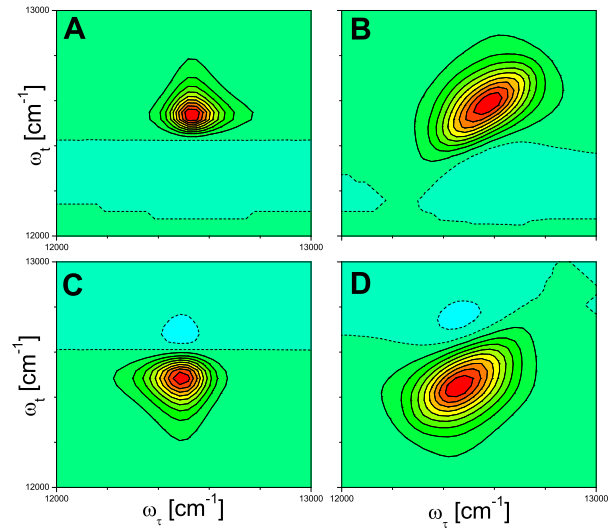}

\caption{\label{fig:Trimer_2D} Two-dimensional correlation photon echo spectrum
of a model trimer. Subfigures: (A) radial configuration $\alpha=-\frac{\pi}{2}$
, single realization; (B) radial configuration, averaged over disorder
with $\Delta=100$ cm$^{-1}$; (C) tangential configuration $\alpha=0$,
single realization; $(D)$ tangential configuration, averaged over
disorder with $\Delta=100$ cm$^{-1}$. All site energies are $\epsilon=12500$
cm$^{-1}$ and $\Gamma=1/300$ fs$^{-1}$. Contours as in Fig. \ref{fig:Dimer-2d}.}

\end{figure}

\begin{figure}
\includegraphics[clip,width=1\columnwidth]{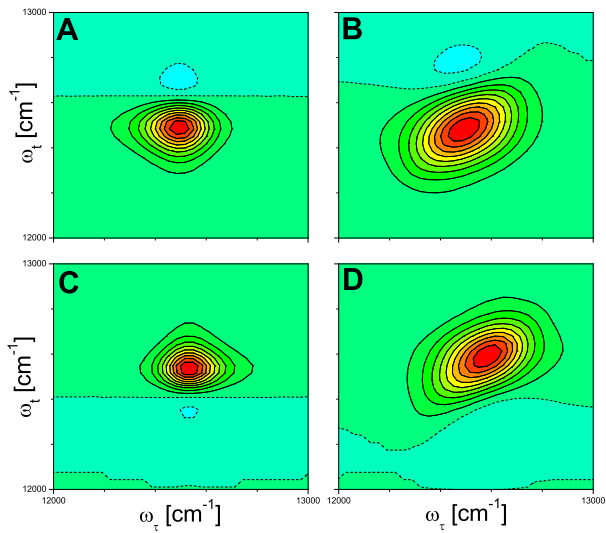}

\caption{\label{fig:Pentamer_2D} Two-dimensional correlation photon echo spectrum
of a model pentamer. Subfigures: (A) radial configuration $\alpha=-\frac{\pi}{2}$
, single realization; (B) radial configuration, averaged over disorder
with $\Delta=100$ cm$^{-1}$; (C) tangential configuration $\alpha=0$,
single realization; $(D)$ tangential configuration, averaged over
disorder with $\Delta=100$ cm$^{-1}$. All site energies are $\epsilon=12500$
cm$^{-1}$ and $\Gamma=1/300$ fs$^{-1}$. Contours as in Fig. \ref{fig:Dimer-2d}.}

\end{figure}

We have also investigated trimers, tetramers and pentamers. As our
work is motivated by highly symmetric homo aggregates like LH2 we
calculated 2D spectra of aggregates of $N$ monomers with circular
$N-$ fold symmetry. Dipole moments are all in plane with the ring
formed by the monomers, and we assume an angle $\alpha$ between the
tangent touching the circle at the position of the monomer and its
transition dipole moment. We compare two cases: $\alpha=0$ (tangential
orientation of the chromophores) and $\alpha=-\frac{\pi}{2}$ (radial
orientation with dipoles pointing towards to center of the ring).
It was shown in Ref. \cite{Read09} that these two configurations
have a distinct position of the excited state absorption. Fig. \ref{fig:Trimer_2D}A
presents 2D spectra of an average trimer with $\alpha=0$, Fig. \ref{fig:Trimer_2D}B
presents the same trimer calculated averaging over $100$ realizations
with energetic disorder of $\Delta=100$ cm$^{-1}$. The former figure
reveals real part of the simple complex Lorentzian lineshape which
is a consequence of the homogeneous limit assumed here. The excited
state absorption is found below the ground state contribution in this
case. For the radial configuration, i.e. $\alpha=-\frac{\pi}{2}$
show on Figs. \ref{fig:Trimer_2D}C and \ref{fig:Trimer_2D}D the
excited state absorption is found above the ground state contribution.
If a trimer is considered a member of the family of $N-$fold symmetric
aggregates this result is the opposite of the expected effect identified
in Ref. \cite{Read09}. However, trimer has to be considered a special
case with respective angles between the chromophores very different
from the larger aggregates of the same symmetry. In larger aggregate
we can expect that the ESA will be in a position similar to the in
line dimer for tangential orientation, and in a position similar to
the sandwich dimer for radial orientation. Indeed, already the pentamer
follows the rule found for larger circular aggregates. As we can see
on Fig. \ref{fig:Pentamer_2D} the two configurations have now position
of the ESA in agreement with Ref. \cite{Read09}. 

Figs. \ref{fig:Dimer-2d} to \ref{fig:Pentamer_2D} demonstrate that
the theory developed in this paper reproduces correctly the ESA features
of 2D spectra of symmetric weakly coupled excitonic aggregates. From
the theoretical point of view ESA features are a result of a delicate
balance between ESA and GSA contributions which cancel exactly in
case of uncoupled chromophores. This feature makes 2D of uncoupled
chromophores additive.

\subsection{Additivity of the 2D spectra}

\begin{figure}
\includegraphics[clip,width=1\columnwidth]{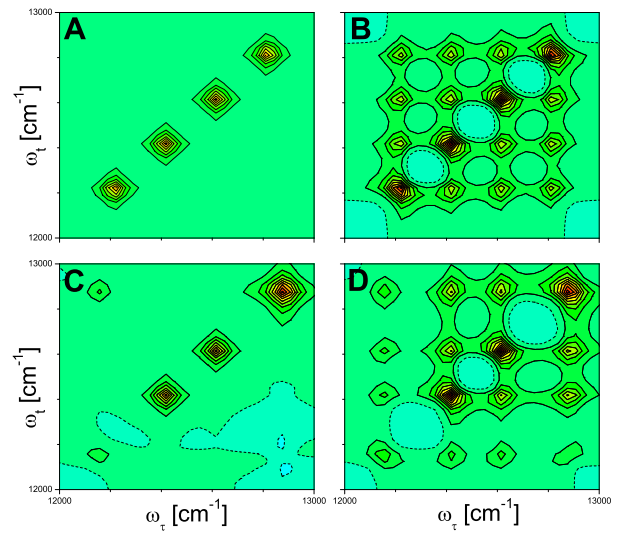}

\caption{\label{fig:tetramer_2D} Demonstration of the additivity of 2D spectra.
Subfigures: (A) Full 2D spectrum of a tetramer consisting of four
uncoupled molecules, (B) the same calculation with ESA ignored, (C)
Tetramer with only the lowest and highest energy molecules coupled
to each other by $J=100$ cm$^{-1}$, (D) The same case with ESA ignored.
Site energies are $\epsilon=12800$, $12600$, $12400$ and $12200$
cm$^{-1}$, disorder width $\Delta=0$ cm$^{-1}$ and $\Gamma=1/150$
fs$^{-1}$. Contours as in Fig. \ref{fig:Dimer-2d}.}

\end{figure}

We will demonstrate that our theory fulfills the additivity property,
and show that the GSA and ESA contributions are correctly balanced.
To this end we consider a hetero tetramer composed of four chromophores
with distinct transition energies $\omega_{1}=12600$ cm$^{-1}$,$\omega_{2}=12500$
cm$^{-1}$,$\omega_{3}=12400$ cm$^{-1}$ and $\omega_{4}=12300$
cm$^{-1}$. Fig. \ref{fig:tetramer_2D}A presents a 2D spectrum of
uncoupled tetramer. No crosspeaks and negative features appear and
the spectrum is a sum of monomeric 2D spectra. When only ground state
to one-exciton band transitions are considered and the aggregate ESA
is ignored many crosspeaks appear in the 2D spectrum (Fig. \ref{fig:tetramer_2D}B).
All these crosspeaks are exactly canceled by the ESA contribution.
When the lowest and highest energy monomers are coupled (here with
$J=200$cm$^{-1}$) the 2D spectrum becomes a sum of two independent
monomers and a coupled dimer, Fig. \ref{fig:tetramer_2D}C. Again,
when the ESA contribution is removed, the redistribution of cross-peak
amplitudes in ground state contribution can be clearly seen, but all
cross-peaks involving the coupled dimer and the independent monomers
are canceled out in Fig. \ref{fig:tetramer_2D}D. 

To calculate 2D spectrum of uncoupled monomers one can therefore simply
calculate individual 2D spectra and sum them. However, the balance
of ESA and the ground state contributions can be disrupted also by
relaxation processes like FRET. Two monomers or two composed systems
that are coupled weakly so that cross-peaks due to mutual interactions
are too weak to be resolved, but nevertheless strong enough to enable
energy transfer will show energy relaxation crosspeaks. This is e.g.
the case of LH3 (see Ref. \cite{Zigmantas06a}). In case of LH3 one
has to account for additional stimulated emission from the states
populated by energy relaxation, and other similar processes. All this
is included in the EM presented in this paper. Moreover, not all states
involved in the theory have to be localized on individual monomers.
We can equally divide the system into parts where excitonic interaction
dominates, and start from excitons formed by this interaction. Mutual
interaction of such blocks is then described by the theory presented
here.

\section{Conclusions\label{sec:Conclusions}}

In this paper we have derived EM for the reduced density matrix of
a system of weakly coupled chromophores interacting with an environment.
The weak excitonic coupling is treated in the second order perturbation
theory and the environmental degrees of freedom are described within
the second cumulant approximation, which for some type of systems
provides an exact solution. We show that our equations are related
to the Förster relaxation rates. In contrast to the usual Förster
type equations, we provide a detailed prescription for the evolution
of coherences. Thus, we are able to describe an effect of dynamic
localization, where the bath destroys not only a wavepacket created
in the complex by ultrafast excitation, but also the coherence established
by the weak resonance coupling. By simulations of model systems in
homogeneous limit we demonstrate that 2D spectroscopy reveals the
excitonic coupling by an offset of the excited state absorption, and
that our local bases description of the systems dynamics is fully
sufficient to account for this effect. 
\begin{acknowledgments}
This work was partially supported by Czech Science Foundation (GACR)
via grant nr. 205/10/0989 and by the Ministry of Education, Youth
and Sports of the Czech republic via grant KONTAKT ME899 and research
plan MSM0021620835. V.B. thanks to the European Physical Society for
the support of his five months stay at Charles University in Prague
in form of University Student Scholarship. L. V. is supported by the
Scientific Council of Lithuania. The spectroscopic package NOSE is
available under GNU Public License at http://www.sourceforge.net.

\appendix
\end{acknowledgments}
\label{sec:Appendices}

\section*{Appendix A: Cumulant expansion evaluation of the $M_{abcd}(t,\tau)$
matrix\label{sec:Appendix-A:-Cummulant}}

Expanding each evolution operator $U_{a}(t)$ up to the second order,\begin{equation}
U_{a}(t)\approx1-i{\cal H}_{a}(t)-{\cal G}_{a}^{(+)}(t),\label{eq:Uexp}\end{equation}
 \begin{equation}
U_{a}^{\dagger}(t)\approx1+i{\cal H}_{a}(t)-{\cal G}_{a}^{(-)}(t),\label{eq:Udegexp}\end{equation}
 where\begin{equation}
{\cal H}_{a}(t)=\frac{1}{\hbar}\int\limits _{0}^{t}d\tau\Delta V_{a}(t),\label{eq:Hdef}\end{equation}
 \begin{equation}
{\cal G}_{a}^{(+)}(t)=\frac{1}{\hbar^{2}}\int\limits _{0}^{t}d\tau\int\limits _{0}^{\tau}d\tau'\Delta V(\tau)\Delta V(\tau'),\label{eq:Gpdef}\end{equation}
 and\begin{equation}
{\cal G}_{a}^{(-)}(t)=\frac{1}{\hbar^{2}}\int\limits _{0}^{t}d\tau\int\limits _{0}^{\tau}d\tau'\Delta V(\tau')\Delta V(\tau),\label{eq:Gmdef}\end{equation}
 the auxiliary matrix $M$ can be evaluated in the second order cummulant
approximation. To that end we evaluate the following traces\begin{equation}
Tr_{Q}\{{\cal H}_{a}(t)W_{eq}\}=0,\label{eq:Tr1}\end{equation}
 \begin{equation}
Tr_{Q}\{{\cal G}_{a}^{(+)}(t)W_{eq}\}=g_{a}(t),\label{eq:Tr2}\end{equation}
 \begin{equation}
Tr_{Q}\{{\cal G}_{a}^{(-)}(t)W_{eq}\}=g_{a}^{*}(t),\label{eq:Tr3}\end{equation}
 \begin{equation}
Tr_{Q}\{{\cal H}_{a}(t){\cal H}_{b}(\tau)W_{eq}\}=g_{ab}(t)-g_{ab}(t-\tau)+g_{ba}^{*}(\tau).\label{eq:Tr4}\end{equation}
 With these results we can evaluate the second order expansion of
$M$ \[
e^{-i\omega_{ab}t-i\omega_{cd}\tau}M_{abcd}(t,\tau)=Tr_{Q}\Big\{\Big[1+i{\cal H}_{a}(t)-{\cal G}_{a}^{(-)}(t)\Big]\]
\[
\times\Big[1-i{\cal H}_{b}(t)-{\cal G}_{b}^{(+)}(t)\Big]\Big[1+i{\cal H}_{c}(\tau)-{\cal G}_{c}^{(-)}(\tau)\Big]\]
 \[
\times\Big[1-i{\cal H}_{d}(\tau)-{\cal G}_{d}^{(+)}(\tau)\Big]W_{eq}\Big\}\]
 \[
=Tr_{Q}\Big\{1+{\cal H}_{a}(t){\cal H}_{b}(t)-{\cal H}_{a}(t){\cal H}_{c}(\tau)+{\cal H}_{a}(t){\cal H}_{d}(\tau)\]
 \[
+{\cal H}_{b}(t){\cal H}_{c}(\tau)-{\cal H}_{b}(t){\cal H}_{d}(\tau)+{\cal H}_{c}(\tau){\cal H}_{d}(\tau)\Big\}\]
 \begin{equation}
-Tr_{Q}\Big\{{\cal G}_{a}^{(-)}(t)+{\cal G}_{b}^{(+)}(t)+{\cal G}_{c}^{(-)}(\tau)+{\cal G}_{d}^{(+)}(\tau)\Big\}.\label{eq:Mexpand}\end{equation}
 Assuming that $g_{ab}(t)=0$ if $a\neq b$ , and taking into account
that $a\neq b$ and $c\neq d$ in the $M$ function we have\begin{equation}
M_{abcd}(t,\tau)=e^{F_{abcd}(t,\tau)+i\omega_{ab}t+i\omega_{cd}\tau},\label{eq:Mabcd_in_g}\end{equation}
 where \[
F_{abcd}(t,\tau)=-g_{a}^{*}(t)-g_{b}(t)-g_{c}^{*}(\tau)-g_{d}(\tau)\]
\[
-\delta_{ac}\left(g_{a}(t)-g_{a}(t-\tau)+g_{a}^{*}(\tau)\right)\]
 \[
+\delta_{ad}\left(g_{a}(t)-g_{a}(t-\tau)+g_{a}^{*}(\tau)\right)\]
\[
+\delta_{bc}\left(g_{b}(t)-g_{b}(t-\tau)+g_{b}^{*}(\tau)\right)\]
 \begin{equation}
-\delta_{bd}\left(g_{b}(t)-g_{b}(t-\tau)+g_{b}^{*}(\tau)\right).\label{eq:Fabcd_in_g}\end{equation}

\section*{Appendix B: Third order response functions\label{sec:Appendix-B:-Third}}

Here we present third order response functions for a three band system
in state representation, using Einstein summation convention. The
upper indices denote the bands so that $g$ corresponds to the ground
state band (for excitonic system only one state is assumed to be in
the ground state band), $e$ and $f$ represent the one-exciton and
two-exciton bands, respectively. Lower indices denote the states within
the bands. In all equations below, index $a$ represents a ground
state index and consequently $a\equiv g$ and $\rho_{a}\equiv\rho_{g}$,
where $\rho_{g}$ is the initial population of the ground state.  \[
R_{1g}(t_{3},t_{2},t_{1})=\langle V_{ji}^{(ge)}V_{gh}^{(eg)}V_{ba}^{(eg)}V_{de}^{(ge)}\rangle\]

\begin{equation}
\times{\cal U}_{ijfh}^{(eg)}(t_{3}){\cal U}_{fgce}^{(ee)}(t_{2}){\cal U}_{cdba}^{(eg)}(t_{1})\rho_{a},\label{eq:R1g_ind}\end{equation}
\[
R_{1f}(t_{3},t_{2},t_{1})=\langle V_{ji}^{(fe)}V_{gh}^{(ef)}V_{ba}^{(eg)}V_{de}^{(ge)}\rangle\]
 \begin{equation}
\times{\cal U}_{ijfh}^{(ef)}(t_{3}){\cal U}_{fgce}^{(ee)}(t_{2}){\cal U}_{cdba}^{(eg)}(t_{1})\rho_{a},\label{eq:R1f_ind}\end{equation}
\[
R_{2g}(t_{3},t_{2},t_{1})=\langle V_{ji}^{(ge)}V_{gh}^{(eg)}V_{ec}^{(eg)}V_{ab}^{(ge)}\rangle\]
 \begin{equation}
\times{\cal U}_{ijfh}^{(eg)}(t_{3}){\cal U}_{fged}^{(ee)}(t_{2}){\cal U}_{cdab}^{(ge)}(t_{1})\rho_{a},\label{eq:R2g_ind}\end{equation}
\[
R_{2f}(t_{3},t_{2},t_{1})=\langle V_{ji}^{(fe)}V_{gh}^{(ef)}V_{ec}^{(eg)}V_{ab}^{(ge)}\rangle\]
 \begin{equation}
\times{\cal U}_{ijfh}^{(ef)}(t_{3}){\cal U}_{fged}^{(ee)}(t_{2}){\cal U}_{cdab}^{(ge)}(t_{1})\rho_{a},\label{eq:R2f_ind}\end{equation}
\[
R_{3g}(t_{3},t_{2},t_{1})=\langle V_{ji}^{(ge)}V_{hf}^{(eg)}V_{ab}^{(ge)}V_{de}^{(eg)}\rangle\]

\begin{equation}
\times{\cal U}_{ijhg}^{(eg)}(t_{3}){\cal U}_{fgce}^{(gg)}(t_{2}){\cal U}_{cdab}^{(ge)}(t_{1})\rho_{a},\label{eq:R3g_ind}\end{equation}
\[
R_{3f}(t_{3},t_{2},t_{1})=\langle V_{ji}^{(fe)}V_{hf}^{(eg)}V_{ab}^{(ge)}V_{de}^{(ef)}\rangle\]
 \begin{equation}
\times{\cal U}_{ijhg}^{(ef)}(t_{3}){\cal U}_{fgce}^{(gf)}(t_{2}){\cal U}_{cdab}^{(ge)}(t_{1})\rho_{a},\label{eq:R3f_ind}\end{equation}
\[
R_{4g}(t_{3},t_{2},t_{1})=\langle V_{ji}^{(ge)}V_{hf}^{(eg)}V_{ec}^{(ge)}V_{ba}^{(eg)}\rangle\]
 \begin{equation}
\times{\cal U}_{ijhg}^{(eg)}(t_{3}){\cal U}_{fged}^{(gg)}(t_{2}){\cal U}_{cdba}^{(eg)}(t_{1})\rho_{a},\label{eq:R4g_ind}\end{equation}
\[
R_{4f}(t_{3},t_{2},t_{1})=\langle V_{ji}^{(ge)}V_{hf}^{(ef)}V_{ec}^{(fe)}V_{ba}^{(eg)}\rangle\]
 \begin{equation}
\times{\cal U}_{ijhg}^{(eg)}(t_{3}){\cal U}_{fged}^{(fg)}(t_{2}){\cal U}_{cdba}^{(eg)}(t_{1})\rho_{a}.\label{eq:R4f_ind}\end{equation}
 The sign $\langle\dots\rangle$ represents orientational averaging
over possible orientations of a molecular system with respect to the
polarization axis of the incident light. The orientational averaging
is preformed for an isotropic distribution of orientations according
to Refs. \cite{HochstrasserAver,MukamelAver}.

\section*{Two-excitation Terms in Equation of Motion}

In this appendix we present the details of ${\cal T}_{2}$ and ${\cal T}_{3}$
terms of the EM, Eq. (\ref{eq:EM_coh2e_1x}), for two-excitation states. 

\[
{\cal T}_{2}=\sum_{c=1}^{K}\Big[\sum_{\gamma=1}^{\sigma-1}\left[R_{ca\pi\gamma}^{*}(t)+R_{\gamma\pi ac}(t)\right]\rho_{c(\gamma\sigma)}(t)\]
\[
+\sum_{\delta=\sigma+1}^{K}\left[R_{ca\pi\delta}^{*}(t)+R_{\delta\pi ac}(t)\right]\rho_{c(\sigma\delta)}(t)\]
 \[
+\sum_{\delta=\pi+1}^{K}\left[R_{ca\sigma\delta}^{*}(t)+R_{\delta\sigma ac}(t)\right]\rho_{c(\pi\delta)}(t)\]
\begin{equation}
+\sum_{\gamma=1}^{\pi-1}\left[R_{ca\sigma\gamma}^{*}(t)+R_{\gamma\sigma ac}(t)\right]\rho_{c(\gamma\pi)}(t)\Big].\label{eq:T2}\end{equation}
\[
{\cal T}_{3}=2\sum_{\alpha=1}^{K}\sum_{\beta=\alpha+1}^{K}\Big(R_{\pi\beta\sigma\alpha}^{*}(t)+R_{\sigma\alpha\pi\beta}^{*}(t)\]
\[
+R_{\pi\alpha\sigma\beta}^{*}(t)+R_{\sigma\beta\pi\alpha}^{*}(t)\Big)\bar{\rho}_{a(\alpha\beta)}(t)\]
\[
+\sum_{\beta=\sigma+1}^{K}\sum_{\delta=\sigma+1}^{K}R_{\pi\delta\delta\beta}^{*}(t)\bar{\rho}_{a(\sigma\beta)}(t)\]
\[
+\sum_{\alpha=1}^{\sigma-1}\sum_{\delta=\sigma+1}^{K}R_{\pi\delta\delta\alpha}^{*}(t)\bar{\rho}_{a(\alpha\sigma)}(t)\]
\[
+\sum_{\beta=\pi+1}^{K}\sum_{\gamma=1}^{\pi-1}R_{\sigma\gamma\gamma\beta}^{*}(t)\bar{\rho}_{a(\pi\beta)}(t)\]
\[
+\sum_{\alpha=1}^{\pi-1}\sum_{\gamma=1}^{\pi-1}R_{\sigma\gamma\gamma\alpha}^{*}(t)\bar{\rho}_{a(\alpha\pi)}(t)\]
\[
+\sum_{\alpha=1}^{\sigma-1}\sum_{\gamma=1}^{\sigma-1}R_{\pi\gamma\gamma\alpha}^{*}(t)\bar{\rho}_{a(\alpha\sigma)}(t)\]
\[
+\sum_{\beta=\sigma+1}^{K}\sum_{\gamma=1}^{\sigma-1}R_{\pi\gamma\gamma\beta}^{*}(t)\bar{\rho}_{a(\sigma\beta)}(t)\]
\[
+\sum_{\delta=\pi+1}^{K}\sum_{\beta=\pi+1}^{K}R_{\sigma\delta\delta\beta}^{*}(t)\bar{\rho}_{a(\pi\beta)}(t)\]
\begin{equation}
+\sum_{\delta=\pi+1}^{N}\sum_{\alpha=1}^{\pi-1}R_{\sigma\delta\delta\alpha}^{*}(t)\bar{\rho}_{a(\alpha\pi)}(t).\label{eq:T3}\end{equation}

\bibliographystyle{prsty}
\bibliography{mancal0215}

\end{document}